\documentstyle[12pt,amssymbols]{article}
\newcommand{\bibiteml}[2]{
                    \bibitem{#1}{#2}
                          }
\newcommand{\calle}[1]{(\ref{#1})}
\begin{document}

\date{}

\title{\Large $V_{(1,1)}^{(t)}$
-PERTURBED MODELS OF CFT\\ AND\\ THEIR QUANTUM GROUP SYMMETRY}

\author{S. A. Apikyan$^\dagger$\\Theoretical Physics Department\\
Yerevan Physics Institute\\ Alikhanyan Br.st.2, Yerevan, 375036 Armenia\\
\\
C. J. Efthimiou$^\ddagger$\\
Newman Laboratory of Nuclear Studies\\Cornell University\\
Ithaca, NY 14853-5001, USA}

\maketitle
\begin{abstract}
We propose a new massive integrable model in quantum field theory.
This model is obtained as a perturbed model of the minimal
conformal field theories
on the hyper-elliptic surfaces by a particular relavant operator
$V_{(1,1)}^{(t)}$. The non-local conserved charges of the model
and their $q$-deformed algebra are also constructed explicitly.
\end{abstract}

\vfill
\hrule
\ \\
\noindent
$\dagger$ e-mail address: apikyan@vx2.yerphi.am    \\
$\ddagger$ e-mail address: costas@hepth.cornell.edu
\newpage

\section{Introduction}

During the last years an essential progress has been
achieved in the investigation
of integrable quantum field theories. Such a success owes much to the fact
that these models are characterized by infinite dimensional Hopf algebra
symmetries, known as affine quantum group symmetries.
These symmetries are genereted by
non-local conserved currents which in many cases
can be constructed explicitly. Such
an approach to the quantum field theory permits to obtain non-perturbative
solutions in the quantum field theory using algebraic methods
\cite{smi}-\cite{bab}.
The situation is analogous to the one taking place in Conformal Field Theory
(CFT). In particular, in CFT, as a result of the
 infinite-dimensional Virasoro algebra
(or other  extended algebras), exact solutions are
successfully obtained  with the help of the  Ward identities \cite{BPZ}.

Explicit currents that generate a $q$-deformation of affine Kac-Moody
algebras \cite{dri},\cite{jim} were constructed for the Sine-Gordon theory and
its generalization to imaginary coupling affine Toda theory in \cite{BL},
and shown to completely characterize the $S$-matrices. At special values
of the coupling where these quantum field theories have ordinary
Lie group $G$ invariance, the quantum affine symmetry becomes the
$G$-Yangian symmetry \cite{ber},\cite{lus}.

The affine quantum group invariance fixes the $S$-matrices up to
overall scalar factors, which in turn  can be fixed
using crossing symmetry,
unitarity and analyticity. These quantum group invariant $S$-matrices,
which are the specializations of the $R$-matrices satisfy the
Yang-Baxter equation.

In the present work a series of new integrable models is identified
and its $q$-deformed structure is studied.
In particular, the organization of the paper is  as follows.
     In section \ref{hyperelliptic-surfaces},
a brief description of the minimal conformal models on
hyper-elliptic surfaces which can be represented as
two-sheet coverings of a ramified sphere is given.
     In section \ref{new-model},
a model of perturbed CFT is proposed; the relevant perturbation
is the highest weight vector of the Virasoro algebra at the
branching points. The characters of this model are calculated
and the existence of an infinite series of  Integrals of Motion (IMs) is
proved; the integrability of the model is thus established.
Furthermore, the  $\beta$-function of the model
is calculated and it is shown that the theory
is massive.
     In the last section, section  \ref{nonlocal-charges},
the non-local currents are constructed. These are related
by non-trivial braiding relations which lead
to the  $q$-deformed
algebra of the conserved charges of the model.

\section{CFT on Hyper-Elliptic Surfaces}
\label{hyperelliptic-surfaces}

Conformal field theories on compact Riemann surfaces, and in particular
on hyper-elliptic surfaces, have been considered by many authors. One
of the pioneering works on hyper-elliptic surfaces was Zamolodchikov's
work
for the Ashkin-Teller models \cite{zam87}; another important contribution
was Knizhnik's work \cite{kni}
 on two-loop
calculations in string theory.  Finally,  in  \cite{CSS},
the minimal models on hyper-elliptic
surfaces were thoroughly  discussed.

Let $\Gamma$ be a compact Riemann surface of genus $g\geq 1$. If $\Gamma$
is a Riemann surface of an algebraic function $y=y(z)$ given by the
equation
\begin{equation}
R(y,z)=y^{n}+a_{1}(z)y^{n-1}+\ldots+a_{n}(z)=0~,
\end{equation}
where $R(y,z)$ is a polynomial of the form shown above,
 then the affine part of $\Gamma$ coincides with
the complex algebraic curve (1,1) in ${\Bbb C}^2$ in case this curve is
ordinary (smooth). Of special importance to us is the example of
hyper-elliptic curves given by equations of the form
\begin{equation}
\label{form1}
y^2=P_{2g+1}(z)~,
\end{equation}
or
\begin{equation}
\label{form2}
y^2=P_{2g+2}(z)~,
\end{equation}
where $P_h(z),~h=2g+1,2g+2,$ is a polynomial of degree $h$ without
roots of multiplity $h$.
In both cases, the
genus of the corresponding Riemann surface is $g$. It is noteworthy that any
Riemann surface of genus $g=1$ or $g=2$ has a representation
in one of the forms  \calle{form1} or \calle{form2},
while the same statement
is not true for surfaces of genus $g=3$. We label the two sheets of the
Riemann surface $\Gamma$
by the numbers $l=0,1$:
\begin{equation}
y^{(l)}(z)= e^{i\pi l}\,P_h^{1/2}(z)=
e^{i\pi l}\,\prod_{i=1}^h\, (z-w_i)^{1/2}~.
\end{equation}
Let $A_a,\, B_a,~a=1,2,\dots,g$ be the basic cycles of the surface.
 As we encircle the point $w_i$  along the
contours $A_a,\, B_a$,
in the case of an $A_a$ cycle  we stay on the
same sheet, while in the case of a
  $B_a$ cycle we  pass from the $l$-th sheet
 to the $(l+1)$-th one.
We shall denote the process of encircling the points $w_i$
on the cycles $A_a, \, B_a$ by the symbols
$\hat{\pi}_{A_a}$, $\hat{\pi}_{B_a}$ respectively.
Here these generators form a group
of monodromy that in our case of two-sheet covering of the sphere coincides
with the ${\Bbb Z}_{2}$ group.

We consider the energy-momentum tensor with representation $T^{(l)}(z)$
on each of these sheets.
The above definition of the monodromy properties
along the cycles $A_a,~B_a$ implies that the following
boundary conditions should be satisfied by the energy-momentum
tensor:
\begin{equation}
\hat{\pi}_{A_{a}}T^{(l)}=T^{(l)} ,\quad \hat{\pi}_{B_{a}}T^{(l)}=T^{(l+1)}~.
\end{equation}
It is convenient to pass to a basis, in which the operators
$\hat{\pi}_{A_a}$, $\hat{\pi}_{B_a}$ are diagonal
\begin{eqnarray}
T=T^{(0)}+T^{(1)}~,&&\quad T^{-}=T^{(0)}-T^{(1)}~,\\
\hat{\pi}_{A_{a}}T=T~,&& \quad \hat{\pi}_{A_{a}}T^{-}=T^{-}~,
\label{BC1}\\
\hat{\pi}_{B_{a}}T=T~,&& \quad \hat{\pi}_{B_{a}}T^{-}=-T^{-}~.
\label{BC2}
\end{eqnarray}
The corresponding operator product expansions
 (OPEs) of the $T,~T^-$ fields can be
determined by taking into account the
 OPEs of $T^{(l)},~T^{(l')}$. On the same
sheet, the OPEs of the two fields $T^{(l)}(z_{1})T^{(l)}(z_{2}),$ are the same
as
that on the sphere, while on different sheets they do not correlate, i.e.
$T^{(l)}(z_{1})T^{(l+1)}(z_{2})\sim {\rm reg}$.
Thus, in the diagonal
basis the OPEs can be found to be
\begin{eqnarray}
T(z_{1})T(z_{2})&=&{c\over 2\,z_{12}^4}+
{2\,T(z_2)\over z_{12}^2}+
{T'(z_2)\over z_{12}} + {\rm reg}~,
\label{OPE1} \\
T^{-}(z_1)T^{-}(z_{2})&=&{c\over 2\,z_{12}^4}+
{2\,T(z_2)\over z_{12}^2}+
{T'(z_2)\over z_{12}} + {\rm reg}~,
\label{OPE2}\\
T(z_1)T^{-}(z_2)&=&{2\over z_{12}^2}\,T^{-}(z_2)+
{T'^{-}(z_2)\over z_{12}}+ {\rm reg}~,
\label{OPE3}
\end{eqnarray}
where $c=2\hat{c}$, and $\hat{c}$ is the central charge in the OPE of
$T^{(l)}(z_{1})T^{(l)}(z_{2})$. It is seen from
\calle{OPE3} that $T^-$ is
primary field with respect to $T$. To write the algebra
\calle{OPE1}-\calle{OPE2} in
the graded form we determine the mode expansion of $T$ and $T^-$:
\begin{eqnarray}
T(z)V_{(k)}(0)&=&\sum_{n\in {\Bbb Z}}\, z^{n-2}L_{-n}V_{(k)}(0)~,\\
T^-(z)V_{(k)}(0)&=&\sum_{n\in {\Bbb Z}}\, z^{n-2-k/2}L_{-n+k/2}^-V_{(k)}(0)~,
\end{eqnarray}
where $k$ ranges over the values 0,1 and determines the parity sector in
conformity with the boundary conditions
\calle{BC1} and \calle{BC2}. Standard calculations
lead to the following algebra for the operators $L_{-n}$ and
$L_{-n+k/2}^{-}$:
\begin{eqnarray}
\lbrack L_n,L_m\rbrack &=& (n-m)\,L_{n+m}+\frac{c}{12}\,
 (n^3-n)\,\delta_{m+n,0}~,\nonumber\\
\lbrack L_{m+k/2}^{-},L_{n+k/2}^{-}\rbrack
 &=&(m-n)\,L_{n+m+k}+\frac{c}{12}\lbrack (m+k/2)^3-
(m+k/2)\rbrack \, \delta_{n+m+k,0}~,~~~~~~
\label{algebra} \\
\lbrack L_m,L_{n+k/2}^- \rbrack &=& \lbrack m-n-k/2\rbrack \, L_{m+n+k/2}^-~.
\nonumber
\end{eqnarray}
The operators $\overline{L}_n,~ \overline{L}_{m+k/2},~\overline{L}_n^-,~
\overline{L}_{m+k/2}^-$ satisfy the same relations and $\overline{L}_n,$
$\overline{L}_{m+k/2},$ $\overline{L}_n^-,$ $\overline{L}_{m+k/2}^-$
commute with $L_n,~
L_{m+k/2},~L_n^-,~L_{m+k/2}^-$.

 To describe
the representations of the algebra \calle{algebra},
 it is necessary to consider
separately the non-twisted  sector  with $k=0$ and
the twisted sector
sector with $k=1$. In order
to write the $\lbrack V_{(k)}\rbrack$ representation of the
algebra \calle{algebra} in a more explicit form, it is convenient to
consider the highest weight states.
In the $k=0$ sector, the
highest weight  state $\vline\, \Delta , \Delta^-\rangle$
is determined with the help of a primary field  $V_{(0)}$
by means of the formula
\begin{equation}
\label{state1}
\vline \,\Delta , \Delta^-\rangle=V_{(0)}\, \vline\, \emptyset
\rangle ~.
\end{equation}
Using the definition of vacuum, it is easy to see that
\begin{equation}
\begin{array}{l}
L_0\,\vline\,\Delta, \Delta^-\rangle=\Delta
\, \vline\, \Delta ,\Delta^-\rangle~ ,\quad
L_0^-\,\vline\, \Delta, \Delta^-\rangle=
\Delta^-\,\vline\, \Delta ,\Delta^-\rangle~, \\
\nonumber\\
L_n\,\vline\, \Delta, \Delta^-\rangle =
L_n^-\,\vline\, \Delta, \Delta^-\rangle=0 ,
\quad n \geq 1~.
\end{array}
\end{equation}
In the $k=1$  sector, we define the vector of highest weight $|\Delta\rangle$
 of the algebra to be
\begin{equation}
\label{state2}
\vline\, \Delta \rangle=V_{(1)}\,\vline \,\emptyset\rangle~,
\end{equation}
where $V_{(1)}$ is a primary field
with respect to $T$. In analogy with the non-twisted sector we
obtain
\begin{equation}
L_0\,\vline \,\Delta \rangle=\Delta \,\vline \,\Delta \rangle,\quad
L_n\,\vline\, \Delta \rangle=L_{n-1/2}^- \,\vline\, \Delta \rangle=0,
\quad n \geq 1~.
\end{equation}
Thus, the Verma module over the algebra \calle{algebra}
 is obtained by the action
of any number of $L_{-m}$ and $L_{-m+k/2}^-$ operators with  $n,m>0$
on the states \calle{state1} and \calle{state2}. As was shown in ref.
\cite{CSS} by means of GKO
(coset construction) method, the central charge of a reducible unitary
representation of the algebra \calle{algebra} has the form
\begin{equation}
\label{ccharge}
c=2-\frac{12}{p(p+1)}=2\hat{c}~ ,\quad p=3,4,\ldots~.
\end{equation}

Using ref. \cite{FF}, Dotsenko and Fateev \cite{DF}
 gave the complete solution for the
minimal model correlation functions on the sphere. They were able to write
down the integral representation for the conformal blocks of
the chiral vertices
in terms of the correlation functions of
the vertex operators of a free bosonic
scalar field $\Phi$ coupled to a background charge $\alpha_0$.
This construction has become known as the Coulomb Gas Formalism
(CGF).
In the present case, this approach is also applicable by
considering  a Coulomb gas for each sheet separately
but coupled to the same bouckground charge:
\begin{equation}
\begin{array}{l}
T^{(l)}=-\frac{1}{4}(\partial_z\Phi^{(l)})^{2} + i\alpha_0\partial_z^2
\Phi^{(l)}~,\quad
\langle\Phi^{(l)}(z)\Phi^{(l')}(z')\rangle=-\delta^{ll'}
\,\ln|z-z'|^2~,\\
\\
\hat{\pi}_{A_a}\partial_z\Phi^{(l)}=\partial_z\Phi^{(l)}~ ,\quad
\hat{\pi}_{B_a}\partial_z\Phi^{(l)}=\partial_z\Phi^{(l+1)}~,
\nonumber
\end{array}
\end{equation}
where $c=2-24\alpha_0^2$ or $\alpha_0^2=1/2p(p+1)$.

Passing to the basis which diagonalizes the operators $\hat{\pi}_{A_a}$ ,
$\hat{\pi}_{B_a}$, i.e.
\begin{eqnarray}
\Phi=\Phi^{(0)} + \Phi^{(1)}~,\quad \Phi^- = \Phi^{(0)} - \Phi^{(1)}
 ~,\nonumber\\
\hat{\pi}_{A_a}\partial_z\Phi = \partial_z\Phi~ ,\quad
\hat{\pi}_{B_a}\partial_z\Phi = \partial_z\Phi~,\\
\hat{\pi}_{A_a}\partial_z\Phi^- = \partial_z\Phi^-~ ,\quad
\hat{\pi}_{B_a}\partial_z\Phi^- = -\partial_z\Phi^-~,
\nonumber
\end{eqnarray}
we finally obtain the bosonization rule for the operators $T$ , $T^-$ in the
diagonal basis
\begin{eqnarray}
T &=& -\frac{1}{4}(\partial_z\Phi)^2 + i\alpha_0\partial_z^2\Phi -
\frac{1}{4}(\partial_z \Phi^-)^2~,\nonumber \\
   \\
T^- &=& -\frac{1}{2}\partial_z\Phi\partial_z\Phi^- +
i\alpha_0\partial_z^2\Phi^- ~.
\nonumber
\end{eqnarray}
In conventions of ref. \cite{CSS},
 the vertex operator with charges $\alpha$, $\beta$
in the $k=0$ (non-twisted) sector is given by
\begin{equation}
\label{vertex1}
V_{\alpha\beta}(z) = e^{i\alpha\Phi+i\beta\Phi^-}~,
\end{equation}
with conformal weights  $\Delta=\alpha^2-2\alpha_0\alpha
+\beta^2$ and $\Delta^-=2\alpha\beta-2\alpha_0\beta$.

In the $k=1$ (twisted) sector the situation is slightly different. Here we
have an antiperiodic bosonic field $\Phi^-$, i.e.
$\Phi^-(e^{2\pi i}z) = -\Phi^-$;
this leads to the deformation of
the geometry of space-time. If we recall that the circle is parametrized by
$\Phi^- \in S^1 \lbrack 0,2\pi R\rbrack$, the
condition  $\Phi^- \sim -\Phi^-$ means that pairs  of points of $S^1$ have been
identified.
Thus, $\Phi^-$ lives on the orbifold $S^1/{\Bbb Z}_2$; under the
identification
 $\Phi^- \sim -\Phi^-$ the two points
$\Phi^-=0$ and $\Phi^-=\frac{1}{2}(2\pi R)$ are fixed points. One can try
to define the twist fields $\sigma_\epsilon(z),~\epsilon=0,1,$ for
the bosonic
field $\Phi^-$, with respect to which $\Phi^-$ is antiperiodic. Notice
that there
is a separate twist field for each fixed point. The OPE of the
current $I^-=i\partial_z\Phi^-$ with the field $\sigma_\epsilon$ is then
\begin{equation}
\begin{array}{l}
I^-(z)\sigma_{\epsilon}(0)=\frac{1}{2}z^{-1/2}\hat{\sigma}_{\epsilon}(0) +
\ldots~,\\
\nonumber\\
I^-(z)\hat{\sigma}_{\epsilon}(0)=\frac{1}{2}z^{-3/2}
\sigma_{\epsilon}(0) + 2z^{-1/2}\sigma'_{\epsilon}(0) + \ldots~.
\end{array}
\end{equation}
The twist fields  $\sigma_\epsilon$ and $\hat{\sigma}_\epsilon$
are primary fields for the $T_{\rm orb}=-\frac{1}{4}(\partial_z\Phi^-)^{2}$
with dimensions $\Delta_{\epsilon}=1/16$ and $\hat{\Delta}_{\epsilon}=
9/16$ respectively. So, in the twisted sector
the highest weight vectors (or primary
fields) can be written as follows
\begin{equation}
\label{vertex2}
V_{\gamma\,\epsilon}^{(t)}=e^{i\gamma\Phi}\sigma_{\epsilon}~ ,\quad
\Delta^{(t)}=\gamma^2-2\alpha_0\gamma+{1\over 16}~.
\end{equation}
In ref. \cite{CSS},
 the anomalous dimensions of the primary fields of the minimal models
for the algebra \calle{algebra}
 were obtained both in the non-twisted and twisted sectors in
conformity with the spectrum of the central charge \calle{ccharge};
in particular, it was found that
the charges $\alpha,\beta,\gamma$
 of the primary fields corresponding to $k=0$ and $k=1$ sectors
have the form:
\begin{equation}
\begin{array}{l}
\alpha_{n'm'}^{nm}={2-n-n'\over 2}\,\alpha_{+}+
{2-m-m'\over 2}\,\alpha_{-}~,\\
\nonumber\\
\beta_{n'm'}^{nm}={n-n'\over 2}\,\alpha_{+}+
{m-m'\over 2}\,\alpha_{-}~,\\
\nonumber\\
\gamma_{nm}={2-n\over 2}\,\alpha_{+}+
{2-m\over 2}\,\alpha_{-}~,\\
\nonumber\\
1\leq n,n'\leq p ,\quad 1\leq m,m'\leq p-1~,
\end{array}
\end{equation}
where the constants $\alpha_{\pm}$
are expressed in terms of the background charge $\alpha_0$:
\begin{equation}
\alpha_{\pm}=\alpha_{0}/2 \pm \sqrt{\alpha_{0}^{2}/4+1/2} ~.
\end{equation}
We denote the corresponding fields by $V^{nm}_{n'm'}$, $V^{(t)}_{nm}$
and their conformal weights by $\Delta^{nm}_{n'm'}$, $\Delta^{(t)}_{nm}$.

We can thus represent the CFT on a hyper-elliptic surface as a CFT on
the plane with an additional symmetry, exactly as
described by the algebra \calle{algebra}.
The corresponding highest weight vectors of the algebra  are
given by \calle{vertex1} and \calle{vertex2}; finally, the
 central charge is given by \calle{ccharge}.

We will confine ourselves to the minimal models on
hyper-elliptic surfaces as presented above; keeping this in mind
 we pass to the construction of
perturbed models of these CFTs.

\section{Perturbation by $V_{nm}^{(t)}$ and Integrals of Motion}
\label{new-model}
\setcounter{equation}{0}

Let $S_p$ be the action the $p$-th conformal  minimal model on the
hyper-elliptic surface $\Gamma$
\begin{equation}
S_p\lbrack\Phi,\Phi^-\rbrack\,\sim \, \int\,d^2z\,(
 \, \partial_z \Phi \partial_{\overline z}\Phi - i\alpha_0R\Phi) +
\int\,d^2z\,\partial_z\Phi^-\partial_{\overline z}\Phi^-~.
\end{equation}
We  now consider the perturbation of this  conformal
field theory by the degenerate relevant operator $V_{nm}^{(t)}$.
\begin{equation}
S_\lambda\,=\,S_p\lbrack\Phi,\Phi^-\rbrack
+\lambda\,\int\,d^2\,z\,e^{i\gamma_{nm}
\Phi(z,\overline{z})}\,\sigma_{\epsilon}(z,\overline{z})~.
\end{equation}
The parameter $\lambda$ is a coupling constant with conformal weight
$(1-\Delta_{nm}^{(t)}\, , \, 1-\Delta_{nm}^{(t)})$.

Obviously, for a generic perturbation the new action $S_\lambda$
does not describe an integrable model.
We are going to choose the perturbation in a way that the corresponding
field theory is integrable.
To prove the integrability of this massive (this claim is proved at the end
of the present section)
theory, one must calculate the characters of the modules
of the identity $I$ and
$V_{nm}^{(t)}$.

The ``basic" currents $T(z)$ and $T^-(z)$ generate an infinite-dimensional
vector subspace $\Lambda$ in the representation space. This subspace can
be constructed by successive applications of the generators $L_{-n}$ and
$L_{-m}^-$ with $n,m>0$ to the  identity operator $I$.
$\Lambda$ can be decomposed to a direct sum of eigenspaces of $L_0$, i.e.
\begin{equation}
\Lambda\,=\,\bigoplus_{s=0}^{\infty} \Lambda_{s}~,\quad
L_0\,\Lambda_s = s\,\Lambda_s~.
\end{equation}
The space $\Lambda$ contains the subspace $\Lambda'=\partial_z\Lambda$.
Therefore, in order to separate the maximal linearly independent set,
one must take
the factor space $\hat{\Lambda}=\Lambda/(L_{-1}\Lambda\,\bigoplus\,L_{-1}^{-}
\Lambda)$ instead of $\Lambda$. The space $\hat{\Lambda}$ admits a similar
decomposition as a direct sum of eigenspaces of $L_0$.
It follows that the formula of the character for $\hat{\Lambda}$
takes the form
\begin{equation}
\chi_0 = (1-q)^2 \prod_{n=1}^{+\infty}\,\frac{1}{(1-q^n)^2}~.
\end{equation}
The dimensionalities of the subspaces
$\hat{\Lambda}_s$ can be determined from the
character formula
\begin{equation}
\sum_{s=0}^{\infty} \, q^s\, \dim(\hat{\Lambda}_s) = (1-q)\,\chi_0 + q~.
\end{equation}
\indent
On the other hand,
the module $V$ of the primary field $V_{nm}^{(t)}$ can be constructed by
successively applying the generators $L_{-k}$ and $L_{1/2-l}^-$
with $k,l>0$ to
the primary field $V_{nm}^{(t)}$. This space $V$ and
the corresponding factor space $\widehat{V} =
V/L_{-1}V$ may also be decomposed in a direct sum of
$L_0$ eigenspaces:
\begin{equation}
V=\bigoplus_{s=0}^{\infty}\,V_s^{(t)}~,\quad L_0\,V_s^{(t)}=s\,
V_s^{(t)}~.
\end{equation}
The dimensionalities of $V_s^{(t)}$
in this factor space associated with the relevant field
\begin{equation}
V_{(1,1)}^{(t)}=e^{i\frac{\alpha_0}{2}\Phi}\sigma_{\epsilon}
\end{equation}
are given by
the character formula
\begin{equation}
\sum_{s={\Bbb N}/2}^{+\infty}\, q^{s+\Delta_{(1,1)}^{(t)}}\,
\dim(\hat{V}_s^{(t)})=
\chi_{\Delta_{(1,1)}^{(t)}}\, (1-q)~,
\label{char1}
\end{equation}
where
\begin{eqnarray}
\label{char2}
\chi_{\Delta_{(1,1)}^{(t)}}&=&q^{\Delta_{(1,1)}^{(t)}}
\prod_{n=1}^{+\infty}\frac{1}{(1-q^{n})(1-q^{n-1/2})}~,\\
\Delta_{(1,1)}^{(t)}&=&\frac{1}{16}\left(1-{6\over p(p+1)}\right)~.
\end{eqnarray}
When the dimensionalities of $\widehat{V}_s^{(t)}$
 (calculated from \calle{char1}, \calle{char2}) are
compared to those of $\hat{\Lambda}_{s+1}$,
we  see that for $s=1,3,5,\dots$  the $\dim(\widehat{\Lambda}_{s+1})$ exceeds
 the $\dim(\widehat{V}^{(t)}_s)$
at least
by the unity, i.e.
$\dim(\widehat{\Lambda}_{s+1})>
 \dim(\widehat{V}^{(t)}_s),~s=1,3,5,\dots~.$
 This proves that the model
\begin{equation}
\label{action}
S_{\lambda}=S_p + \lambda\,\int\,d^2z\,e^{i\frac{\alpha_0}{2}
\Phi(z,\overline{z})}\,\sigma_{\epsilon'}(z,\overline{z})
\end{equation}
possesses an infinite set of non-trivial IMs.
We
note here that there are no such IMs for
perturbations by the operators $V_{nm}^{(t)}$ with $n,m>1$.

We now  briefly study the renormalization
group flow behaviour in the vicinity of the fixed point.

Solving the Callan-Symanzik equation \cite{IZ} up to third order, one can
obtain the $\beta$-function
\begin{equation}
\beta=\varepsilon\, g\, \left( 1 + \frac{Y}{6}\,  g^2\right) + {\cal O}(g^4) ~.
\end{equation}
In the above equation, we have denoted
\begin{equation}
\varepsilon = 1-\Delta_{(1,1)}^{(t)}
\end{equation}
and
\begin{equation}
Y = \int d^2 z_1 \int d^2 z_2 \,\langle V_{(1,1)}^{(t)}(z_1,\overline{z}_1)
V_{(1,1)}^{(t)}(z_2,\overline{z}_2)V_{(1,1)}^{(t)}(1,1)
V_{(1,1)}^{(t)}(0,0) \rangle ~.
\end{equation}
Since $Y>0$, we conclude  that there is no reason
to expect the exsistance of any non-trivial zeros of the $\beta$-function.
In the absence of
 zeros, the field theory described by the action \calle{action}
 has a finite correlation length
$R_c\sim \lambda^{-1/2\varepsilon}$ and the spectrum consists of particles
with non-zero mass
of order $m\sim R_c^{-1}$. In this case, the IMs force the scattering
of the particles to be  factorizable, i.e. there is particle production,
the set of particle momenta is preserved, the $n$-particle
$S$-matrix is a product of 2-particle $S$-matrices  etc.

\section{Infinite Quantum Group Symmetry}
\label{nonlocal-charges}
\setcounter{equation}{0}

\indent
In this section we briefly review the method developed in ref.
\cite{BL} and then we apply it to our model.

We consider a CFT perturbed by a relevant operator with zero Lorentz spin.
The Euclidean action is given by
\begin{equation}
\label{pert-action}
S_\lambda=S_{\rm CFT}+\frac{\lambda}{2\pi}
\,\int\,d^2z\,V_{\rm pert}(z,\overline{z})~,
\end{equation}
where the perturbation field can be written as $V_{\rm pert}(z,\overline{z})=
V_{\rm pert}(z)\overline{V}_{\rm pert}(\overline{z})$
 (or a sum of such terms but
in our case this is irrelevant).
Let us assume that for the conformal
invariant action $S_{\rm CFT}$ there exist the chiral currents $J(z)$,
$\overline{J}(\overline{z})$ satisfying equations $\partial_{\overline
z}J(z)=0$,
$\partial_z\overline{J}(\overline{z})=0$. Then for the  action
\calle{pert-action} $S_\lambda$, the
perturbed currents, which are local with respect to the perturbing field, up
to the first order, are given by Zamolodchikov's equations \cite{zam89}
\begin{equation}
\begin{array}{l}
\partial_{\overline z}J(z,\overline{z})=\lambda\oint_z\,
{d\omega\over 2\pi i}\, V_{\rm pert}
(\omega,\overline{z})J(z)~,\\ \\
\partial_z\overline{J}(z,\overline{z})=\lambda\oint_{\overline{z}}\,
{d\overline{\omega}\over 2\pi i}\,
V_{\rm pert}(z,\overline{\omega})\overline{J}(\overline{z})~.
\end{array}
\end{equation}
The condition for the conservation of the currents up to first order in
perturbation theory is that the residues of OPEs appearing in the above
contour integrals are total derivatives:
\begin{equation}
\begin{array}{l}
{\rm Res}\Big(V_{\rm pert}(\omega)J(z)\Big)=\partial_zh(z)~,
\\  \\
{\rm Res}\Big(\overline{V}_{\rm pert}(\overline{\omega})
\overline{J}(\overline{z})\Big)
=\partial_{\overline{z}}
\overline{h}(\overline{z})~.
\end{array}
\end{equation}
Then Zamolodchikov's equations for the currents are written in the form
\begin{equation}
\label{continuity-equation}
\begin{array}{l}
\partial_{\overline{z}}J(z,\overline{z})=\partial_zH(z,\overline{z})~,\\
\\
\partial_z\overline{J}(z,\overline{z})=
\partial_{\overline{z}}\overline{H}(z,\overline{z})~,
\end{array}
\end{equation}
where the fields $H$, $\overline{H}$ are
\begin{equation}
\begin{array}{l}
H(z,\overline{z})=\lambda\, \lbrack h(z)\overline{V}_{\rm pert}(\overline{z})
+\dots\rbrack~,\\  \\
\overline{H}(z,\overline{z})=\lambda\,\lbrack
V_{\rm pert}(z)\overline{h}(\overline{z})+\dots\rbrack~,
\end{array}
\end{equation}
where the dots represent contributions coming from terms in the OPEs
which are more singular
 than the residue term.
The conserved charges following from the conserved currents
\calle{continuity-equation}
 are
\begin{equation}
\label{charges}
\begin{array}{l}
Q=\int\,{dz\over 2\pi i}\,J+\int {d\overline{z}\over 2\pi i}\, H~,\\ \\
\overline{Q}=\int\,{d\overline{z}\over 2\pi i}\,\overline{J}
+\int\,{dz\over 2\pi i}\,\overline{H}~.
\end{array}
\end{equation}
Using the non-trivial braiding relations between the
conserved currents, one can
obtain the $q$-deformed affine Lie algebra for the conserved charges
\calle{charges}.

We are now going to implement the above construction of non-local charges
for  the theory described by the action \calle{action}.
We will thus derive the $q$-deformed Lie algebra underlying the theory.
Using the construction explained above, we can show that the
action \calle{action} admits the following non-local conseved
quantum currents:
\begin{equation}
\label{continuity2}
\begin{array}{l}
\partial_{\overline{z}}J =\partial_zH~,\\
\nonumber\\
\partial_z\overline{J}=\partial_{\overline z}
\overline{H}~,
\end{array}
\end{equation}
where
\begin{equation}
\label{currents}
\begin{array}{l}
J=\colon e^{ia\varphi(z)}\,
e^{ib\varphi^-(z)}\colon\, \sigma(z)~,\\  \\
\overline{J}=
\colon e^{ia\overline{\varphi}(\overline{z})}e^{ib
\overline{\varphi}^-(\overline{z})}\colon
\,\overline{\sigma}(\overline{z})~,\\  \\
H(z,\overline{z})=\lambda\, A \, \colon
e^{i(a+\alpha_0/2) \varphi (z)}e^{i(b+k) \varphi^-(z)}
\overline{\sigma}(\overline{z})
e^{i\frac{\alpha_{0}}{2}
\overline{\varphi}(\overline{z})}\colon~,\\ \\
\overline{H}(z,\overline{z})=\lambda\, A\,
\colon e^{i(a+\alpha_0/2) \overline{\varphi}(\overline{z})}
e^{i(b+k) \overline{\varphi}^-(\overline{z})}
\sigma (z)e^{i\frac{\alpha_0}{2}\varphi(z)}
\colon~,
\end{array}
\end{equation}
and
\begin{eqnarray}
a &=& -(15/8+k^{2})/(\alpha_{0}+4k^{2}/\alpha_{0})~,
\nonumber\\
b &=& 2k a/\alpha_0~,
\label{constants}\\
A &=& \alpha_0/2(a + \alpha_0/2)~.\nonumber
\end{eqnarray}
In the derivation of \calle{currents}, we used the OPEs
\begin{equation}
\begin{array}{l}
\sigma(z)\, \sigma(x)=(z-x)^
{k^2-1/8}:e^{ik\varphi^{-}(x)}:+\ldots~,\\ \\
\overline{\sigma}(\overline z)\overline{\sigma}(\overline x)=
(\bar z-\bar x)^{\overline{k}^2-1/8}
\,:e^{i\overline{k}\overline{\varphi}^-(\overline{x})}:+\ldots~.
\end{array}
\end{equation}
 From the continuity equations
\calle{continuity2} we define the conserved charges
\begin{equation}
\begin{array}{l}
Q =\int\,\frac{dz}{2\pi i}\,J + \int\,\frac{d\overline{z}}{2\pi i}\,H
 ~,\\
\nonumber\\
\overline{Q} =\int\,\frac{dz}{2\pi i}\,\overline{H} +
\int\frac{d\overline{z}}{2\pi i}\,\overline{J}~.
\end{array}
\end{equation}
To find the commutation relations between the charges $Q$ and
$\overline{Q}$, we must first  derive the braiding relations
of the non-local
conserved currents $J$, $\overline{J}$. To this end we will make
use of the well known identity
\begin{equation}
e^A\,e^B=e^B\,e^A\,e^{\lbrack A,B\rbrack}~,
\quad  \lbrack A,\lbrack A, B\rbrack\rbrack=
\lbrack B,\lbrack A,B\rbrack\rbrack=0~.
\end{equation}
We then obtain the following braiding relations
\begin{equation}
\begin{array}{ll}
e^{ia\varphi(z)}e^{ib\varphi(z')}=
e^{\mp i\pi ab}\,e^{ib\varphi(z')}e^{ia\varphi(z)}~,
 &\quad z\lessgtr z'~,\\  \\
e^{ia\varphi^{-}(z)}e^{ib\varphi^{-}(z')}=
e^{\mp i\pi ab}\,e^{ib\varphi^-(z')}e^{ia\varphi^-(z)}
 ~,  &\quad z\lessgtr z'~,\\ \\
e^{ia\overline{\varphi}(\overline{z})}
e^{ib\overline{\varphi}(\overline{z}')}=
e^{\pm i\pi ab}\,e^{ib\overline{\varphi}(\overline{z}')}
e^{ia\overline{\varphi}(\overline{z})}~,
 &\quad \overline{z}\lessgtr \overline{z}'~,\\ \\
e^{ia\overline{\varphi}^-(\overline{z})}
e^{ib\overline{\varphi}^-(\overline{z}')}=
e^{\pm i\pi ab}\,e^{ib\overline{\varphi}^-(\overline{z}')}
e^{ia\overline{\varphi}^-
(\overline{z})}~,  &\quad \overline{z}\lessgtr \overline{z}'~,\\ \\
e^{ia\varphi(z)}e^{ib\overline{\varphi}(\overline{z}')}=e^{i\pi ab}
\,e^{ib\overline{\varphi}(\overline{z}')}
e^{ia\varphi(z)}~, &\quad \forall z,\overline{z}'~,\\ \\
e^{ia\varphi^-(z)}e^{ib\overline{\varphi}^-(\overline{z}')}=
e^{i\pi ab}\,e^{ib\overline{\varphi}^-(\overline{z}')}e^{ia\varphi^-(z)}~,
 &\quad \forall z,\overline{z}'~.
\end{array}
\end{equation}
Using the representation  of the twist fields $\sigma,
\overline{\sigma}$ in terms of
scalar bosonic fields which was proposed in ref. \cite{AZ},
we can derive the following
braiding relations:
\begin{equation}
\begin{array}{ll}
\sigma(z)\sigma(z')=e^{\mp i\pi/8}\,\sigma(z')\sigma(z)~,&\quad z\lessgtr z'~,
\\   \\
\overline{\sigma}(\overline{z})\overline{\sigma}
(\overline{z}')=e^{\pm i\pi/8}\,\overline{\sigma}(\overline{z}')
\overline{\sigma}(\overline{z})~,&\quad \overline{z}\lessgtr \overline{z}'~,
\\ \\
\sigma(z)\overline{\sigma}(\overline{z}')=
e^{+i\pi/8}\,\overline{\sigma}(\overline{z}')\sigma(z)~,&
\quad \forall z,\overline{z}'~.
\nonumber
\end{array}
\end{equation}
Consequently the non-local conserved currents have the non-trivial braiding
relations
\begin{equation}
J(x,t)\overline{J}(y,t)=
q^{\nu}\,\overline{J}(y,t)J(x,t)~,
\end{equation}
where
\begin{equation}
q=e^{-i\pi}~,\quad
\nu = 1/8-aa-bb~.
\end{equation}
\indent
Using the above braiding relations and the expressions
\calle{currents}, one finds that
the conserved charges satisfy the relations
\begin{eqnarray}
Q\overline{Q}-q^{\nu}\,\overline{Q}Q
=\frac{\lambda}{2\pi i}\,\int_t\,
(dz\partial_z+d\overline{z}\partial_{\overline{z}})\,
A\, e^{i(a+\alpha_0/2)\varphi(z)}
e^{i(b+k)\varphi^{-}(z)}\times \nonumber\\
\times A\,e^{i(a+\alpha_{0}/2)\overline{\varphi}(\overline{z})}
e^{i(b+k)\overline{\varphi}^-(\overline{z})}~.
\label{QQ}
\end{eqnarray}

Now let us recall that the scalar field $\varphi^-$ lives on the orbifold
$S^1 / {\Bbb Z}_2$ and hence the momentum $k$ must be
quantized. Therefore, the above
relations must be transformed to
\begin{eqnarray}
\widehat Q_{\epsilon}\widehat{\overline{Q}}_{\overline{\epsilon}}-
q^{\nu_{\epsilon\overline{\epsilon}}}\,
\widehat{\overline{Q}}_{\overline{\epsilon}}\widehat Q_{\epsilon}&=&
{\lambda\over 2\pi i}\, \sum\,
A_L^{nm}A_R^{nm}\, \int_t \, (dz\,\partial_z+
d\overline{z}\,\partial_{\overline{z}})\times \nonumber\\
&\times & e^{i(a_L^{nm}+\alpha_0/2)\varphi(z)+
i(a_R^{nm}+\alpha_0/2)\overline{\varphi}(\overline{z})}\times\nonumber\\
&\times & e^{i(b_L^{nm}+k_L^{nm})\varphi^-(z)+
i(b_R^{nm}+k_R^{nm})\overline{\varphi}^-(\overline{z})}~,
\end{eqnarray}
where
\begin{equation}
\begin{array}{l}
\nu_{\overline{\epsilon}\epsilon}=
1/8-a_L^{nm}a_R^{nm}-b_L^{nm}b_R^{nm}
 \\ \\
k_L^{nm}=k_L^{nm}(\epsilon,\epsilon')=
{n\over R} + \left( m+{\epsilon+\epsilon'\over 2}
\right)\,{R\over 2}~,
\\ \\
k_R^{nm}=k_R^{nm}(\overline{\epsilon},\epsilon')=
{n\over R}-\left( m+
{\overline{\epsilon}+\epsilon'\over 2}\right) \,{R\over 2}~.
\nonumber
\end{array}
\end{equation}
The constants $a_L^{nm}$, $a_R^{nm}$, $b_L^{nm}$,
$b_R^{nm}$, $A_L^{nm}$, $A_R^{nm}$ are obtained from the
relations \calle{constants} and
$\epsilon,\overline{\epsilon},
\epsilon'\in\{0,1\}$.

Finally,  the topological charge for the model \calle{action}
is defined as follows:
\begin{eqnarray}
{\cal T}_{\rm top}&=&\int_{-\infty}^{+\infty}\,dx\,\partial_x\Phi(x)+
\int_{-\infty}^{+\infty}\,dx\,\partial_x\Phi^-(x)\nonumber\\
&=&\int_{-\infty}^{+\infty}\,dx\,\partial_x\,(\varphi +
 \overline{\varphi})+
\int_{-\infty}^{+\infty}\,dx\,\partial_x(\varphi^- +
\overline{\varphi}^-)
\nonumber\\
&=&T_{\rm top}+\overline{T}_{\rm top}+
T_{\rm top}^-+\overline{T}_{\rm top}^-~,
\label{top-charg}
\end{eqnarray}
where $\Phi$, $\Phi^-$ and the quasi-chiral components
$\varphi, \overline{\varphi},
\varphi^-,\overline{\varphi}^-$ are related by the following
equations:
\begin{equation}
\begin{array}{l}
\varphi(x,t)=\frac{1}{2}\,
\left(\Phi(x,t)+\int_{-\infty}^x\, dy\, \partial_t
\Phi(y,t)\right)~,\\
\nonumber\\
\overline{\varphi}(x,t)=\frac{1}{2}\,\left(\Phi(x,t)-
\int_{-\infty}^x\, dy\,\partial_t
\Phi(y,t)\right)~,\\
\nonumber\\
\varphi^-(x,t)=\frac{1}{2}\,\left(\Phi^-(x,t)+
\int_{-\infty}^x\,dy\,\partial_t\Phi^-(y,t)\right)~,\\
\nonumber\\
\overline{\varphi}^-(x,t)=\frac{1}{2}\,\left(\Phi^-(x,t)-
\int_{-\infty}^x\,dy\,
\partial_t\Phi^-(y,t)\right)~,
\end{array}
\end{equation}
These equations guarantee that
 $\Phi=\varphi+\overline{\varphi}$ and $\Phi^-=\varphi^-+
\overline{\varphi}^-$. Taking into account all these, the
right hand side of the equation \calle{QQ}
can be reexpressed in terms of the usual topological charges
charge in \calle{top-charg}:
\begin{eqnarray}
\widehat{Q}_\epsilon\widehat{\overline{Q}}_{\overline{\epsilon}} -
q^{\nu_{\overline{\epsilon}\epsilon}}\,
\widehat{\overline{Q}}_{\overline{\epsilon}}\widehat{Q}_\epsilon
= \frac{\lambda}{2\pi i}\,
\sum\, A_L^{nm}A_R^{nm}\,
\lbrack 1-e^{i(a_L^{nm}+\alpha_0/2)T_{\rm top}+
i(a_R^{nm}+\alpha_0/2)\overline{T}_{\rm top}}\times\nonumber\\
\times e^{i(b_L^{nm}+k_L^{nm})T_{\rm top}^-+
i(b_R^{nm}+k_R^{nm})\overline{T}_{\rm top}^-}\rbrack~.~~~~~
\label{QQ2}
\end{eqnarray}
Then, one can easily calculate the commutators
\begin{equation}
\label{TQ}
\begin{array}{l}
\lbrack T_{\rm top},Q_\epsilon^{nm}\rbrack=
 a_L^{nm}\, Q_{\epsilon}^{nm}~,\quad
\lbrack \overline{T}_{\rm top},
\overline{Q}_{\overline{\epsilon}}^{nm}\rbrack=
a_R^{nm}\,\overline{Q}_{\overline{\epsilon}}^{nm}~, \\ \\
\lbrack T_{\rm top}^-,Q_{\epsilon}^{nm}\rbrack=
b_{L}^{nm}\, Q_{\epsilon}^{nm}~,\quad
\lbrack\overline{T}_{\rm top}^-,
\overline{Q}_{\overline{\epsilon}}^{nm}\rbrack=
b_R^{nm}\,\overline{Q}_{\overline{\epsilon}}^{nm}~.
\end{array}
\end{equation}
Thus, these commutation relations \calle{TQ}
together with the relations \calle{QQ2}
constitute the
algebra, to the lowest non-trival order in perturbation theory,
which is the symmetry of the $S$-matrix of the theory.

Unfortunately, the isomorphism between the algebra
\calle{QQ2},\calle{TQ} and the Hopf
algebra has not been established yet, and, hence, the universal $R$-matrix of
this hidden Hopf algebra has not been studied. However, we are going to make
some additional comments about these open questions in the near future.

\section{Conclusions}

To summarize, in the present paper we have introduced a new
integrable  model
in quantum field theory. The novelty of the model resides in the fact that
it is built on a hyper-elliptic surface instead of the usual
Euclidean plane. The quantum symmetry of the model has been identified
in terms of the non-local conserved charges.  This has led to a generalization
of the method first introduced by Bernard and LeClair \cite{BL}
for the affine Toda field theories where only boson fields are involved.
As is understood very well by now, the quantum non-local conserved charges
provide a quantum field theoretic basis for understanding quantum groups.
Unfortunately, the mapping from the physical algebra satisfied by the non-local
charges to the $q$-deformed Lie algebra has not been discovered yet.
If this mapping is found, one will be able to study the universal $R$-matrix
and consequently uncover the structure of the $S$-matrix.

\vspace{.5cm}
{\bf Acknowledgements}

We would like to thank A. LeClair, F. Smirnov and R. Poghossian for helpful
discussions.

\end{document}